\documentclass[12pt]{article}
\usepackage{epsfig,amsmath,amsfonts}
\usepackage[figuresright]{rotating}
\newcommand{\bm}[1]{\mbox{\boldmath $#1$}}
\begin{document}
\title{Statistics of temperature fluctuations in a buoyancy dominated
boundary layer flow simulated by a Large-eddy simulation model}
\author{Marta Antonelli$^1$, Andrea Mazzino$^{2,1}$ and Umberto Rizza$^{2}$\\
\small{$^1$ INFM--Department of Physics, University of Genova, I--16146
Genova, Italy}\\
\small{$^2$ ISAC/CNR - Sezione di Lecce - Strada provinciale 
Lecce-Monteroni km 1.2 73100 Lecce}}
\date{\today}
\maketitle
\begin{abstract}
Temperature fluctuations in an  atmospheric convective boundary layer
are investigated by means of Large Eddy Simulations (LES).
A novel statistical characterization for both weak temperature fluctuations
and strong  temperature fluctuations has been found. Despite the
nontriviality of the dynamics of temperature fluctuations, our data
support the idea that the most relevant statistical properties can be
captured solely in terms of two scaling exponents, 
characterizing the entire mixed layer.
Such exponents control asymptotic (i.e. core and tails) 
rescaling properties of the probability density functions of 
equal-time
temperature differences, $\Delta_r \theta$,  between points separated 
by a distance ${\bm r}$. A link between statistical properties of
large temperature fluctuations and geometrical properties of the
set hosting such fluctuations is also provided.
Finally, a possible application of our new findings to the problem 
of subgrid-scale parameterizations for the temperature field
in a convective boundary layer is discussed.  
\end{abstract}

\vspace{4mm}

\section{Introduction}
Temperature in an atmospheric boundary layer (ABL) is typically convected by
a velocity background, ${\bm v}$, and diffuses by the action of
molecular motion and/or small-scale turbulent eddies. The basic
equation governing such a process is the well-known
advection-diffusion equation (Pielke, 1984) for the (potential)
temperature, $\theta$,
\begin{equation}
\partial_t\theta + v_{\alpha} \partial_{\alpha}\theta=D_0\partial^2\theta +
S_{\theta}\;\;\;,
\label{FP}
\end{equation}
where $S_{\theta}$ represents the sources and sinks of heat, 
eventually present within the domain, ${\bm v}$ is the 
velocity field advecting the
temperature, and $D_0$ may represent either the diffusion
coefficients
or, alternatively, an eddy diffusion coefficient if one intends to
focus on the large scale behavior (i.e.~large eddies) of $\theta$
and thus needs to parameterize, 
in some way, small scale temperature dynamics. Repeated indexes are summed.\\
Here, we use the following short notations:
$\partial_t\equiv \partial\bullet/\partial t$;  
$\partial_i\equiv \partial\bullet/\partial x_i$, $i=1,\cdots ,3$;
$\partial^2\equiv\sum_{i=1}^3\partial_i\partial_i\bullet$.

In many situations, temperature dynamics, driven by the
velocity field via the advection term in Eq.~(\ref{FP}), does not
react back on the velocity field. This is the case, for instance,
in a neutrally stratified boundary layer (occurring, e.g., 
under windy conditions with a complete cloud cover) where buoyancy forces are
negligible compared
to the other terms in the Navier-Stokes (NS) equations ruling the velocity
field dynamics. In this case, temperature behaves as a passive scalar 
as a good approximation. \\
As a matter of fact, during the diurnal cycle  
neutral stratification is rarely observed in the ABL (Garratt, 1999).
More frequently, one observes stable conditions (occurring, e.g., at night in
response to surface cooling by longwave emission to space) 
or unstable, convective, ones (occurring, e.g., when strong surface heating
produces thermal instability or convection).
The role of temperature is active in both cases, that amounts to say
that temperature drives the velocity field dynamics through the buoyancy
contribution (usually modeled by means of the well-known Boussinesq
coupling). The latter contribution is now the leading one in the NS equations.

The main feature characterizing both active and passive scalar
dynamics is the presence of strong fluctuations in the temperature
field. Such fluctuations affect the whole range of scales involved in
the temperature dynamics, from the largest scales of motion
to the smallest ones where diffusive effects become important.
This huge excitation of degrees of freedom,
gives meaning to the term ``scalar
turbulence'' recently used to denote the dynamics
of temperature fluctuations (Shraiman and Siggia, 2000).

The first unpleasant consequence of persistent fluctuations is the failure
of  any attempt to construct dimensional theories for the statistics 
of temperature fluctuations (Frisch, 1995), quantitatively defined as 
temperature differences $\Delta_r\theta$ between points separated by $r$:
$\Delta_r\theta\equiv\theta({\bm r},t)-\theta({\bm 0},t)$. 
The common strategy
for dimensional approaches consists to define {\it typical}
length/time scales 
and {\it typical} amplitude for the fluctuations of
the unknown fields (e.g., $\Delta_r \theta$ and $\Delta_r{\bm v}$)
and then try to balance the various terms in the basic equations
(e.g.~Eq.~(\ref{FP}) coupled to the NS equations)
to deduce predictions for $\Delta_r \theta$ and 
$\Delta_r{\bm v}$ as a function of the separation ${\bm r}$.\\
This is, for instance, the essence of the first dimensional
theory for scalar turbulence due to Kolmogorov, Obukhov and Corrsin
(1949) and Bolgiano (1959). As a result of these theories,
probability density functions (pdfs), $P(\Delta_r\theta)$, 
of temperature differences, $\Delta_r \theta$, obey the following simple 
rescaling: $P(\Delta_r\theta)=r^{-\alpha}\tilde{P}(\Delta_r\theta/r^{\alpha})$,
where $\tilde{P}$ is a function of the sole ratio 
$\Delta_r\theta/r^{\alpha}$.\\
Such property immediately implies that, dimensionally, one has
$\Delta_r\theta\sim r^{\alpha}$ and, for the $p$-th moment of $\Delta_r\theta$:
$\langle(\Delta_r \theta)^p\rangle\sim r^{\zeta_p}$, with $\zeta_p$
(known as scaling exponents) linear function of $p$:  $\zeta_p=\alpha p$\\
The linearity of $\zeta_p$ vs $p$ reflects the fact that only one parameter,
$\alpha$, is necessary to explain most of the statistical properties of
$\theta$.
One has, in other words, `single-scale fluctuations';
that amounts to say that it is irrelevant which part of the
probability density function of temperature differences is sampled to
define a typical fluctuation.

Rather than the above simple scenario predicted by dimensional
theories, turbulent systems show an infinite hierarchy  of
`independent' fluctuations (Frisch, 1995), that amounts to say the
strong dependence on the order $p$ considered to define the 
typical fluctuation. More quantitatively, turbulent systems exhibit 
a nonlinear behavior of $\zeta_p$ vs $p$\footnote{Such curve must be
concave and not decreasing,
as it follows from general inequalities of probability theory
(see, e.g.,~ Frisch (1995)).} where the infinite set of exponents, 
$\zeta_p$, select different classes of fluctuations. The departure of the
actual curve of $\zeta_p$ vs $p$ from the linear (dimensional)
prediction is named ``intermittency'' or ``anomalous scaling''
(Frisch, 1995). Intermittency is probably the most representative
property characterizing a turbulent system.

Our aim here is to provide a statistical characterization of
temperature fluctuations in a convective boundary layer dominated by
well-organized plumes, simulated by
a large eddy simulation model (Moeng, 1984).\\
We shall focus, in particular, on the statistical characterization 
of two different classes of fluctuations: weak temperature
fluctuations, mainly occurring in the inner plume regions and strong
temperature fluctuations, associated to the plume interfaces (see Fig.~1). 
As we shall show, weak fluctuations are associated to linear behavior
of scaling exponents $\zeta_p$ vs $p$ for small $p$'s, 
while strong fluctuations
(captured by large $p$'s) cause the so-called intermittency saturations, i.e.
the curve $\zeta_p$ vs $p$ tends a to a constant value, $\zeta_{\infty}$,  
for $p$ large enough. The saturation exponent $\zeta_{\infty}$  is
simply connected (see Sec.~\ref{geom}) 
to the fractal dimension of the set hosting 
large temperature excursions: $\zeta_{\infty}=d-D_F$, where $d$ 
and  $D_F$ are the  usual dimension of the space and the fractal 
dimension of the large temperature fluctuation set, respectively.
\\
Despite the complexity of temperature fluctuation dynamics in a 
convective boundary layer (CBL), 
reflected in the strong intermittency of the system,
only two exponents are necessary to capture most of the statistics
of temperature fluctuations.  \\
It is worth emphasizing that the same statistical 
characterization has been recently found 
for two dimensional idealized models of both passive 
(see Frisch et al (1999), Celani et al
(2000) and Celani et al (2001a)) 
and active scalar turbulence (see Celani et al (2001b))  
simulated by means of both direct numerical
simulations and Lagrangian methods (see, e.g., Frisch et al (1998)).
This points toward the generality of our new findings
within the context of scalar transport.

\section{Statistical tools}
\label{tools}
The aim of this section is to provide a quick summary of
the statistical tools we have exploited to characterize temperature 
fluctuations in a CBL. \\
The basic and well-known indicator is the probability density function,
$P(\Delta_{{\bm r}; {\bm x}}\theta)$, 
of temperature differences, $\Delta_{{\bm r}; {\bm x}}\theta$, 
over a scale $r$, 
defined as:
\begin{equation}
\Delta_{{\bm r}; {\bm x}}\theta \equiv \theta({\bm x}+
{\bm r},t)-\theta({\bm x},t)\;\;\; .
\label{SF} 
\end{equation}
The pdf $P(\Delta_{{\bm r}; {\bm x}}\theta)$ 
will depend other than on ${\bm r}$
also on ${\bm x}$ if the system is not homogeneous. In our CBL,
we have homogeneity along $x$-$y$ planes, but not
along the vertical direction $z$. We thus shall have a dependence on the 
vertical coordinate $z$.
Moreover, in the analysis of Sec.~\ref{risu}, separations ${\bm r}$ will be
taken along $x$-$y$ planes in the direction forming an angle
of $\pi /4$ with the geostrophic wind direction. We shall thus denote
our pdf simply as $P(\Delta_{r; z}\theta)$. The choice for the
direction $\pi /4$ has been done in order to
reduce the contamination of the
scaling exponents by anisotropic effects.
More details on this `magic' angle can be found in 
Celani et al. (2001a).

By definition, weak fluctuations (i.e. small values of $|\theta({\bm x}+
{\bm r},t)-\theta({\bm x},t)|$ with respect to a typical fluctuation
defined  as 
$\sigma^{(z)}\equiv [\langle\theta^2\rangle- \langle\theta\rangle^2
]^{1/2}$) 
are associated to the pdf core while, on the contrary, large
fluctuations
(i.e. $|\theta({\bm x}+
{\bm r},t)-\theta({\bm x},t)| \gg \sigma^{(z)}$) are associated to the 
pdf tails.\\
The above considerations can be easily paraphrased 
in terms of the moments, 
$S_p({\bm r};z)\equiv \langle(\Delta_{r;z} \theta)^p\rangle$,
known as structure functions.
Large fluctuations are captured by large $p$'s, while weak fluctuations
by small $p$'s.

Let us now  introduce two possible behaviors for the pdf 
$P(\Delta_{r; z}\theta)$. As we shall show in the sequel,
such behaviors will characterize, within the entire mixed layer,
the statistical properties of
weak temperature fluctuations and strong temperature fluctuations, 
respectively.\\

\noindent{\bf Self-similar behavior}

In terms of probability density functions of
$\Delta_{r;z}\theta$,
such a  behavior is defined by the rescaling property:
\begin{equation}
P(\Delta_{r;z}\theta)=r^{-\alpha^{(z)}}\tilde{P}
\left (\frac{\Delta_{r;z}\theta}{r^{\alpha^{(z)}}}\right )\;\;\; .
\label{pdf-resc1}
\end{equation}
It can be immediately verified from the definition of moments:
\begin{equation}
S_p({\bm r};z)=\int_{-\infty}^{+\infty} P(\Delta_{r;z}\theta)
(\Delta_{r;z}\theta)^p d(\Delta_{r;z}\theta)
\label{mom}
\end{equation}
that (\ref{pdf-resc1}) is equivalent to the following behavior 
for the structure functions, $S_p$:
\begin{equation}
S_p({\bm r};z)\sim r^{\zeta_{p}^{(z)}}\ \ \mbox{with}\ \ \ 
\zeta_{p}^{(z)}=\alpha^{(z)} p, ~~~ \alpha^{(z)}>0\;\;\; ,
\label{lin} 
\end{equation}
that is a linear behavior, with the factor $\alpha^{(z)}$ in general depending
on the elevation $z$ within the mixed layer.

It is worth noticing that (\ref{pdf-resc1}) and (\ref{lin})
does not necessarily imply a Gaussian shape for
$P(\Delta_{r;z}\theta)$. On the contrary, if $P(\Delta_{r;z}\theta)$
is Gaussian then (\ref{lin}) (and thus (\ref{pdf-resc1})) are
immediately satisfied as it follows from the well-known property of 
Gaussian statistics (see, e.g., Frisch (1995)): 
$S_{2p}({\bm r};z)=(2p-1)!!\;[S_2({\bm r};z)]^p$,
together with the assumption that $S_2({\bm r};z)\sim r^{2\alpha^{(z)}}$.\\

\noindent{\bf Intermittency saturation}
(i.e.~the strongest violation of dimensional predictions):

In terms of $P(\Delta_{r;z}\theta)$, intermittency saturation
is defined as 
\begin{equation}
P(\Delta_{r;z}\theta)=\frac{r^{\zeta_{\infty}^{(z)}}}{\sigma^{(z)}}
Q\left (\frac{\Delta_{r;z}\theta}{\sigma^{(z)}}\right )
~~~\mbox{for}~~~|\Delta_{r;z}\theta|>\lambda\sigma^{(z)}~~~(\lambda >1)\;\;\; ,
\label{resca}
\end{equation}
where $Q$ is some function (not determined a {\it priori}) which 
does not depend on the separation $r$. \\
In terms of cumulated probabilities, i.e.~the sum (integral)
of the pdfs over the large temperature fluctuations (i.e.~for
$|\Delta_{r;z}\theta| > \lambda \sigma^{(z)}$, with $\lambda>1$),
defined as:
\begin{equation}
Prob[|\Delta_{r;z}\theta|>\lambda\sigma^{(z)}]
\equiv \int_{-\infty}^{-\lambda\sigma^{(z)}}
P(\Delta_{r;z}\theta) d(\Delta_{r;z}\theta)+
\int_{\lambda\sigma^{(z)}}^{+\infty}
P(\Delta_{r;z}\theta) d(\Delta_{r;z}\theta),
\label{cum}
\end{equation}
saturation is equivalent to the following power law behavior, holding 
for different values of $\lambda>1$:
\begin{equation}
Prob[|\Delta_{r;z}\theta|>\lambda\sigma^{(z)}]
\sim r^{\zeta_{\infty}^{(z)}}\;\;\; .
\label{pawer}
\end{equation}
The scaling exponents, $\zeta_{\infty}^{(z)}$, can be thus easily extracted
by measuring the slope of 
$\log \left\{Prob[|\Delta_{r;z}\theta|>\lambda\sigma^{(z)}]\right\}$ vs
$\log r$.

Finally, it is worth observing that,
in terms of structure functions intermittency saturations means:
\begin{equation}
S_p({\bm r};z)\sim r^{\zeta_{p}^{(z)}}\ \ \mbox{with}\ \ \ 
\zeta_{p}^{(z)}=\zeta_{\infty}^{(z)}, ~~~
\mbox{for}~~~p>p_{crit}\;\;\; ,
\label{sat} 
\end{equation}
as one can easily verify from the definitions of moments (\ref{mom})
and from (\ref{resca}).\\
The scaling exponents $\zeta_p^{(z)}$ thus tend to 
a constant value $\zeta_{\infty}^{(z)}$
for orders, $p$'s large enough. Such behavior justifies the
word `saturation' to denote the laws (\ref{resca}) and (\ref{pawer}).

\section{The Large-Eddy simulation model}

In order to gather statistical informations on the turbulent structure
of a CBL, we used the LES code described in Moeng (1984)
and Sullivan et al (1994). Such model 
has been widely used and tested to investigate fundamental 
problems in the framework of boundary layers
(see, e.g., Moeng and Wyngaard (1989), Moeng et al. (1992), 
Andr\'en and Moeng (1993), Moeng and Sullivan (1994), among the
others).\\ For this reason we confine ourselves only on 
general aspects of the LES strategy. Details can be found
in the aforementioned references.

The key point of the LES strategy is that the large scale motion 
(i.e. motion associated to the large turbulent eddies) is explicitly solved 
while the smallest scales (typically in the inertial range of scales)
are described in a statistical consistent way
(i.e. parameterized in terms of the resolved, large scale, 
velocity and temperature fields).
This is done by filtering the governing equations for velocity and
potential temperature
by means of a filter operator. Applied, e.g., to the  potential
temperature field $\theta$, the filter
is defined as the convolution:
\begin{equation}
\overline{\theta}({\bm x})=\int \theta({\bm x}')G({\bm x}-{\bm x}')d{\bm x}'
\end{equation}
where $\overline{\theta}$ is the filtered variable and $G({\bf x})$ is a
tridimensional filter function. The field $\theta$ can be
thus decomposed as 
\begin{eqnarray}
\label{decom}
\theta=\overline{\theta}+{\theta}' .
\end{eqnarray}
Applying the filter operator both to the Navier--Stokes equations 
and to the equation for the potential temperature, and exploiting 
the decomposition (\ref{decom}) (and the analogous for the velocity field)
in the advection terms one 
obtains the corresponding filtered equations. For the sake of brevity,
we report the sole filtered equation for the potential temperature:
\begin{eqnarray}
\label{filtNS2} 
\partial_t\overline{\theta}=-\overline{\overline{v}_{\alpha}\partial_{\alpha}
\overline{\theta}}-\partial_{\alpha}\tau^{(\theta)}_{\alpha}
\end{eqnarray}
where $\tau^{(\theta)}_{\alpha}$ are the subgrid turbulence fluxes of
virtual temperature (in short SGS fluxes). They are related to the
resolved-scale field
as 
\begin{equation}
\tau^{(\theta)}_{\alpha}=-K_{H}\partial_{\alpha}\overline{\theta}
\end{equation}
$K_{H}$ being the SGS eddy coefficient for heat. 
A similar expression holds for the subgrid turbulence fluxes of
momentum (see Moeng, 1984) that are defined in terms of the 
SGS eddy coefficient for momentum ($K_{M}$).\\
The above two eddy coefficients are related to the velocity scale
$\overline{e'}^{1/2}$, $\overline{e'}$ being the SGS turbulence energy the 
equation of which is solved in this model, and to the length scale
$l\equiv (\Delta x\Delta y\Delta z )^{1/3}$ (valid for the convective
cases) $\Delta x$, $\Delta y$, and $ \Delta z $ being the grid mesh
spacing in $x$, $y$ and $z$. Namely:
\begin{equation}
K_{M}=0.1\; l\; \overline{e'}^{1/2}
\end{equation}
\begin{equation}
K_{H}=3 K_{M}.  
\end{equation}
 
\section{The simulated  convective experiment}
\label{simula}

In the present first study, our attention has been focused on the
Simulation B (hereafter referred to as Sim B) by Moeng and Sullivan
(1994). Sim B is a buoyancy-dominated flow with a 
relatively small shear effect, where vigorous thermals (see again Fig.~1) 
set up due to buoyancy force.\\
The sole difference of our simulation
with respect to the Moeng and Sullivan's simulation 
is the increased spatial resolution, here of $128^3$ grid points.\\
A preliminary sensitivity test at the lower resolution 
$96^3$ (as in  Moeng and Sullivan (1994)) did not show
significant differences in the results we are going to present.
Sensitivity tests at higher resolutions are still in progress
and seem to confirm our preceding conclusion.

Our choice for a convective boundary layer lies on the fact that, 
in such regimes,
dependence of resolved fields
on SGS parameterization should be very weak, and thus LES strategy
appears completely justified.
Indeed, in  convective regimes,  SGS motion acts as net 
energy sinks that drain energy from
the resolved motion. This is another way to say that energy blows
from large scales of motion toward the smallest scales and 
the cumulative (statistical) effect of the latter scales can be 
successfully captured
by means of simple eddy-diffusivity/viscosity SGS models. 
Uncertainties
eventually present at the smallest scales directly affected by 
SGS parameterizations (that are not the concern of the analysis
we are going to show)
do not propagate upward but are promptly diffuse
(and thus dissipated)  owing to the action of the aforementioned 
eddy-diffusivity/viscosity character of SGS motion. 
Genuine inertial range dynamics can thus develop and, as we shall see,
the typical features
characterizing an inertial range of scales (e.g., rescaling properties
of statistical objects) to appear. 

The following parameters characterize
the Sim B. Geostrophic wind, $U_g=10\;m/s$;
friction velocity, $u_*= 0.56\; m/s$; convective
velocity, $w_*= 2.02\; m/s$;
PBL height, $z_i= 1030\;m$; large-eddy
turnover time, $\tau_*= 510\;s$; stability parameter, $z_i/L=-18 $ ($L$
being the Monin--Obukov length);
potential temperature flux at the surface, $Q_*=0.24\;m K/s$.\\
Moreover, the numerical domain size in the $x$, $y$ and $z$ directions
are $L_x=L_y=5\;km$ and $L_z=2\;km$, respectively; the time step for the
numerical integration is about $1\;s$. 
For details on  the simulated experiment, readers can
refer to Moeng and Sullivan (1994).

To perform our statistical analysis, we first reached the
quasi-steady state. It took, as in Moeng and Sullivan (1994), 
about six large-eddy turnover times, $\tau_*$.
After that time, a new simulation has been made for about $37 \tau_*$ 
and the simulated potential temperature field saved at $0.5 \tau_*$
intervals for the analysis. Our data set was thus formed by 74
(almost independent)  potential temperature snapshots.

Each simulation hour required about 24 computer hours on an
Alpha-XP1000  workstation. 

\section{Results and discussions}
\label{risu}
\subsection{Statistics of large temperature fluctuations}
Let us start our statistical analysis from the large temperature
fluctuations. These are controlled by the pdf tails of temperature
differences, $\Delta_{r;z}\theta$, and, as we are going to show, they
are compatible with
the laws (\ref{resca}) and (\ref{pawer}), that means
intermittency saturation. \\
To show that, it is enough to see whether or not there exist a
positive number, $\zeta_{\infty}^{(z)}$, such that the quantities
$\sigma^{(z)} P(\Delta_{r;z}\theta) r^{-\zeta_{\infty}^{(z)}}$ 
collapse on the
same curve, $Q$, for different values of the separation $r$. Indeed, as
showed in Sec.~\ref{tools}, in the presence of saturation
the function $Q$, appearing in (\ref{resca}),
does not depend on $r$. \\
The validity of (\ref{resca}) can be seen in Fig.~2, where the
behavior of $P(\Delta_{r;z}\theta)$ for $z/z_i=0.3$ and two 
values of $r$ are shown, 
$z$ and $z_i$ being the elevation above the bottom boundary and
the mixed layer height, respectively. In the graph (a), 
$P(\Delta_{r;z}\theta)$ is reported without any $r$-dependent 
rescaling; in (b) 
we show $\sigma^{(z)}P(\Delta_{r;z}\theta) r^{-\zeta_{\infty}^{(z)}}$ for
$\zeta_{\infty}^{(z)}\sim 0.6$. The data collapse occurring on the 
tails of the curves of graph (b) is the footprint of intermittency 
saturation.\\ 
In Figs.~3 and 4 we show the analogous of Fig.~2 but for 
$z/z_i=0.45 $ and $z/z_i=0.6 $. Also in these cases, the exponent
giving the data collapse is $\zeta_{\infty}^{(z)}\sim 0.6$.
Similar behaviors have been observed for all $z$'s within the mixed layer.\\
As a conclusion, from the evidences of Figs.~2, 3 and 4, it turns out that
the saturation exponent $\zeta_{\infty}^{(z)}$ does not depend on $z$
within the mixed layer. It is thus a property of the entire mixed
layer and, for this reason, it will be simply denoted by
$\zeta_{\infty}$.

Let us now corroborate the scenario of intermittency saturation
by looking at the cumulated probability (\ref{cum}). \\
For the saturation to occur,  such probability 
has to behave as a power
law with exponent  $\zeta_{\infty}$ (see (\ref{pawer})).  
Such behavior is indeed observed and showed 
in Fig.~4(a) (for $z/z_i=0.3$ and $\lambda = 5$ and $5.5$)
and in Fig.~4(b) (for $z/z_i=0.6$ and $\lambda = 5$ and $5.5$).
The continuous lines have the slope $\zeta_{\infty}\sim 0.6$ as measured from
Figs.~2 and 3. The fact that there exist a region of scales, $r$, where
that slope is parallel to the slope of the cumulated probabilities
means, again, intermittency saturation with a unique
(i.e.~characterizing the whole mixed layer) exponent.\\
It is worth noting 
that figures similar to Figs.~5(a) and 5(b) have been obtained
also for smaller values of $\lambda$, e.g., $\lambda=2.5$ and
$\lambda=3.5$. Population of strong events 
being decreasing as $\lambda$ increases, the above independence on
$\lambda$  points for the robustness of our statistics.

In Fig.~6(a) we report, for $z/z_i=0.45$, the behaviors of the 
sixth and eight-order structure functions of temperature
differences 
vs the separation between points (squares). Stright lines have the
slope $\zeta_{\infty}=0.6$.  This is a further, direct, evidence of
intermittency saturation. To investigate the statistical convergence
of our sixth and eight-order  moments, we reported in  Fig.~6(b) 
the bulk contribution to such moments: 
$(\Delta_r \theta)^p P(\Delta_r \theta)$,
with $p=6$ and $p=8$,
$r/L\sim 7\times 10^{-2}$ in the inertial range of scales. 
For comparison $P(\Delta_r \theta)$
is also shown. Note that the maximum contribution
to the moments six and eight comes  from fluctuations $\Delta_r \theta$
in the region where $P(\Delta_r \theta)$ (i.e., $p=0$) is well
resolved, i.e., our statistics appears reliable up to the order eight.

Once $\zeta_{\infty}$ is known, we evaluated from (\ref{resca})
the unknown function $Q$. Such function is
shown in Fig.~7 for two different values of $z$ within the mixed layer. 
Differences among the two curves are evident, signaling that $Q$
contains a dependence on the elevation $z$. Such dependence can be
associated to the relatively small shear  present in our
convective simulation (see Sec.~\ref{simula}) .\\
Further simulations spanning intermediate ABL (i.e.~where both shear
and buoyancy are important) have however to be performed in order
to confirm the above last conclusion. 

It is worth stressing that the scales $r$ at which we observe 
scaling behaviors are always larger than $\sim 8$ grid-points
(i.e.~sufficiently far from the scales directly affected
by SGS parameterizations). 
Our attention being focused in a region sufficiently 
far from boundaries, this is another point in favor for the
possible SGS independence of our results.

\subsubsection{A link between geometry and statistics}
\label{geom}
Let us now conclude this section with a geometric point of view 
for the intermittency saturation (see also Celani et al, 2001a).
As we shall see, the saturation exponent $\zeta_{\infty}$ is related
to the fractal dimension, $D_F$, of the set hosting the strong
temperature fluctuations. \\
To do that, let us schematize in a very rough way our strong 
(i.e.~larger than some $\sigma^{(z)}$)
temperature fluctuations in the form of quasi-discontinuities 
(i.e.~step functions). Each quasi-discontinuity 
will define a point (we are performing the analysis on 
planes at constant $z$) of given coordinate  
in our two-dimensional plane. The ensemble of all
points defines the set, $S$, hosting strong temperature fluctuations.
Roughly, $S$ is formed by the intersection of our two-dimensional
plane with the plume interfaces across which
strong temperature jumps occur.\\
A useful indicator to characterize geometrically our set, $S$, is the 
fractal dimension, $D_F$, (see, e.g., Frisch (1995) for a presentation 
oriented toward turbulence problems). We briefly recall the standard
way to define $D_F$.\\
\begin{itemize}
\item Take boxes of side, $r$, and cover the whole plane at fixed $z$.
Denote with $N_{tot}$ the total number of those boxes;
\item Define the function $N(r)$ as the number of boxes containing
at least one point of $S$;
\item For $r$ sufficiently small, one expects power law behavior for
$N(r)$ in the form: $N(r)\sim r^{-D_F}$, which defines the fractal
dimension of $S$.  
\end{itemize}
Given the fractal dimension, $D_F$, it is now easy to compute
the probability, $Prob[|\Delta_{r;z}\theta|>\lambda\sigma^{(z)}]$,
of having strong (i.e.~larger than some $\sigma^{(z)}$)
temperature jumps within a certain
distance $r$. Indeed, by definition, we have:
\begin{equation}
Prob[|\Delta_{r;z}\theta|>\lambda\sigma^{(z)}]
\equiv\frac{\mbox{favorable cases}}{\mbox{possible cases}}=
\frac{N(r)}{N_{tot}}\sim \frac{r^{-D_F}}{r^{-2}}=r^{2-D_F}\;\;\; .
\label{frac}
\end{equation}
From (\ref{pawer}) and (\ref{sat}) the identification
$\zeta_{\infty}=2-D_{F}$
immediately follows. Notice that if one does not restrict the
attention on the sole planes at constant $z$, but focuses  on the whole
three dimensional space, the above relation becomes
$\zeta_{\infty}=3-D_{F}'$ where $D_{F}'$ is the fractal dimension
of the new set $S$.

\subsection{Statistics of weak temperature fluctuations}
Let us now pass to investigate the statistics of well-mixed regions of
the temperature field, corresponding to the inner parts of plumes that
are likely to be present in our CBL (see again Fig.~1).\\
In these regions, fluctuations turn out to be very gentle and,
as an immediate consequence, statistics is expected to be controlled
in terms of single-scale fluctuations (see the Introduction).
The best candidate to characterize, from a statistical point of view,
weak fluctuations is thus the rescaling form given by (\ref{pdf-resc1}). \\
To investigate whether or not our data are compatible with such
rescaling, it is enough to verify
whether there exist a number, $\alpha^{(z)}$, ({\it a priori} dependent
on the elevation $z$ within the mixed layer) such that, 
looking at $r^{\alpha^{(z)}}P(\Delta_{r;z}\theta)$ vs
$\Delta_{r;z}\theta/r^{\alpha^{(z)}}$ for different values of $r$, all
curves collapse one on the other for each value of $z$.\\
Our data support this behavior for the pdf cores (as expected, the rescaling
(\ref{pdf-resc1}) holds solely for weak fluctuations), as it can be 
observed in Fig.~2(c) (for $z/z_i=0.3$), in Fig.~3(c) 
(for $z/z_i=0.45$) and Fig.~4(c)  (for $z/z_i=0.6$). 
In all cases, the values of $\alpha^{(z)}$
is $\sim 0.2$, that means that $\alpha^{(z)}$ does not depend on $z$.
As the exponent $\zeta_{\infty}$, $\alpha\equiv\alpha^{(z)}$ thus 
characterizes the entire mixed layer.

\section{Conclusions and discussions}
We have characterized, from a statistical point of view, both large and weak 
temperature fluctuations of a convective boundary layer simulated by a 
large eddy simulation model. \\
The main results of our study can be summarized as follows.
\begin{itemize}
\item Large temperature fluctuations, occurring across plume interfaces,
turn out to be strongly intermittent. This is the cause of the observed 
break down of mean field theories {\'a} la Kolmogorov, predicting a linear 
behavior of the scaling exponents, $\zeta_p$, of  
the structure functions of temperature differences, vs the order $p$. 
We found, on the contrary,
a pdf rescaling which corresponds to 
a nonlinear shape of $\zeta_p$ vs $p$, with 
$\zeta_p\to\zeta_{\infty}\equiv const$ for $p$ large enough. This behavior
is named {\it intermittency saturation}, i.e.~the strongest violation of
dimensional predictions.\\
Hence, the concept of `typical fluctuation'
does not make sense: it is necessary to specify which part of the pdf of 
temperature differences is sampled for the definition of `typical fluctuation'.

\item Weak temperature fluctuations, characterizing the inner plume region
where temperature  is extremely well-mixed,  
have a self-similar character. This amounts to say 
that, despite the fact that many scales are excited in the well-mixed regions,
the concept of `typical fluctuation' here makes sense. In this case
a simple rescaling characterizes the pdf core, which corresponds to
a linear behavior of the curve $\zeta_p$ vs $p$ for small $p$'s.
The slope of the straight line $\zeta_p$ vs $p$ 
is $\alpha\sim 0.2$.

\item Exponents $\alpha$ and $\zeta_{\infty}$ appear to be independent on 
the elevation within the mixed layer. They are thus an intrinsic property of 
the entire mixed layer.

\item Statistics and geometry turn out to be intimately related.
A simple relationship holds indeed between $\zeta_{\infty}$ and the 
fractal dimension, $D_F$, of the set hosting the large temperature 
fluctuations:
$\zeta_{\infty}=d-D_F$ where $d$ is the usual dimension of the physical 
space.\\
As for $\zeta_{\infty}$, $D_F$ appears an intrinsic, i.e.~$z$-independent,
property of the entire mixed layer.
\end{itemize}

It is worth stressing that the present scenario holds also for
idealized, two-dimensional, models of scalar turbulence both passive
(see Frisch et al (1999), Celani et al
(2000) and Celani et al (2001a)) 
and active (see Celani et al (2001b)),  
simulated by means of direct numerical simulations.
This fact naturally points toward the  possible generality of the 
present statistical characterization for the entire class of scalar 
transport problems.

Finally, let us discuss a possible application of our results
within  boundary layer physics, and, more specifically, in the
LES approach. As well known, one of the most challenging
problem in the  LES strategy is to find a proper way to describe
the dynamical effect of small-scale unresolved motion on the
resolved large scale dynamics.
Recently, new approaches have emerged as alternatives to the eddy
viscosity and similarity models (see, e.g., Meneveau and Katz, 2000).
They construct the small-scale unresolved part of a total
field (e.g., the velocity field) by extrapolating properties
of the (resolved) coarse-grained field. A specific
form for the subgrid scale field is thus postulated 
exploiting scale-invariance (i.e.~inertial range scaling behavior)
of the coarse-grained field.
Note that, standard approaches postulate the form of the stress
tensor rather than the structure of the unresolved field.\\
The mathematical tool which permits to perform such an interpolation
is known as ``fractal interpolation'' (see Meneveau and Katz
(2000) and references therein) where the free parameter of the
theory is the fractal dimension of the field.\\
Up to now, many efforts have been devoted to exploit such
strategy for the velocity fields. The latter exhibit however 
a  multifractal structure (roughly speaking,
 an infinite set of fractal dimensions characterizes the whole field)
and thus the fractal dimension parameter is a sort of `mean field'
description. \\
The suggestion arising from
our results is that the same technique exploited for the velocity field
appears  successfully applicable
in the convective case to the temperature field as well. Indeed,
our results support a fractal structure of the temperature
field. The situation seems to be even better than that 
for the velocity field. As we have stressed in the preceding
sections, solely two exponents characterize most of the statistical
properties of temperature fluctuations. We thus propose to schematize
temperature fluctuations as a bi-fractal object (i.e., the simplest
multifractal object) described by our two exponents $\alpha$ and 
$\zeta_{\infty}$, and to generalize ``fractal interpolation'' to 
this case.\\
To do that, it seems necessary to investigate how the exponents
we found for the analyzed experiment change by varying, e.g., the
weight of buoyancy with respect to shear. We are currently working
on this point and, as far as we remain on convective experiments,
it seems that only small variations appear. \\
It should also be  interesting to investigate what happens to the
above scenario in stable stratified boundary layers. In that
cases LES approach appears more delicate than in a CBL.   
Observations in field and/or in a wind tunnel 
should  be thus necessary to investigate the problem.

\section*{References}

\noindent  Andr\'en, A., and C.-H. Moeng, 1993: 
Single-point closure in a neutrally 
stratified boundary layer. {\it J. Atmos. Sci.}, {\bf 50}, 3366-3379. \\

\noindent  Bolgiano, R., 1959: Turbulent spectra in a stably 
stratified atmosphere.  {\it J. Geophys. Res.}, {\bf 64}, 2226-2229.\\

\noindent  Celani, A., A. Lanotte, A. Mazzino, and M. Vergassola, 2000:
Universality and saturation of intermittency
in passive scalar turbulence.
{\it Phys. Rev. Lett.}, {\bf 84},  2385-2388.\\

\noindent  Celani, A., A. Lanotte, A. Mazzino, and M. Vergassola, 2001a:
Fronts in passive scalar turbulence.
{\it Phys. Fluids}, {\bf 13}, 1768-1783.\\

\noindent  Celani, A., A. Mazzino, and M. Vergassola, 2001b:
Thermal plume turbulence.
{\it Phys. Fluids}, {\bf 13}, 2133-2135.\\

\noindent Frisch, U., 1995: Turbulence. {\it The legacy of A.N. 
Kolmogorov.} Cambridge University  Press, 296 pp.\\

\noindent Frisch, U., A. Mazzino, and M. Vergassola, 1998:
Intermittency in passive scalar advection.
{\it Phys. Rev. Lett.}, {\bf 80}, 5532-5537.\\

\noindent Frisch, U., A. Mazzino, and M. Vergassola, 1999:
Lagrangian dynamics and high-order moments
intermittency in passive scalar advection.
{\it Phys. Chem. Earth}, {\bf 24}, 945-951.\\

\noindent Garratt, J.R., 1999: {\it The atmospheric boundary layer.} 
Cambridge University Press, 316 pp.\\

\noindent Meneveau, C. and J. Katz, 2000: 
Scale-Invariance and turbulence models for Large-eddy simulation,
{\it Annu. Rev. Fluid Mech.}, {\bf 32}, 1-32.\\

\noindent Moeng, C.-H., 1984: A large-eddy-simulation model for the study of 
planetary boundary-layer turbulence, {\it J. Atmos. Sci.}, {\bf 41}, 2052-2062. \\

\noindent Moeng, C.-H. and J.C. Wyngaard, 1989: Evaluation of turbulent transport and dissipation closures in second-order modeling. 
{\it J. Atmos. Sci.}, {\bf 46}, 2311-2330. \\

\noindent Moeng C.-H., and P.P. Sullivan, 1994: A comparison of shear and 
buoyancy driven Planetary Boundary Layer flows.
{\it J. Atmos. Sci.}, {\bf 51}, 999-1021.\\

\noindent Obukhov, A., 1949: Structure of the temperature field in turbulence. 
{\it Izv. Akad. Nauk. SSSR. Ser. Geogr.}, {\bf 13}, 55-69.\\

\noindent Pielke, R.A., 1984: Mesoscale Meteorological Modeling.
Academic Press, 612 pp.\\

\noindent Shraiman B.I, and E.D. Siggia, 2000: Scalar turbulence. {\it
Nature},
{\bf 405}, 639-646.\\

\noindent Sullivan, P.P., J.C.~McWilliams, and C.-H.~Moeng, 1994:
A subgrid-scale model for large-eddy simulation of planetary boundary layer
flows. {\it Bound. Layer Meteorol.}, {\bf 71}, 247-276.\\

\newpage

\vskip 0.2cm
\noindent {\bf Acknowledgments}\\
We are particularly grateful to Chin-Hoh Moeng  and Peter Sullivan,
for providing us with their LES code as well as
many useful comments, discussions and suggestions. 
Helpful discussions and suggestions by
A.~Celani, R.~Festa, C.F.~Ratto and M.~Vergassola are also acknowledged.
This work has been partially supported by the INFM project GEPAIGG01
and Cofin 2001, prot. 2001023848.
Simulations have been performed at CINECA (INFM parallel computing initiative).

\newpage

\section*{List of Figures}

\begin{enumerate}

\item 
A typical snapshot of the potential temperature field $\theta$,
in the quasi-steady state of a convective boundary layer
simulated by a Large Eddy Simulation with resolution $128^3$.
Above: vertical cross-section restricted to the mixed
layer; below: horizontal  cross-section inside the  mixed
layer. Colors are coded
according to the intensity of the field: white corresponds to large
temperature, black to small ones. Plumes and well-mixed regions are
clearly detectable.

\item 
The pdf's $P(\Delta_{r;z} \theta)$, for two values of $r$
inside the inertial range of scales (dotted lines: $r/L=0.22$; 
continuous line: $r/L=0.11$, $L$ being the side of the (squared)
simulation domain) and $z/z_i=0.3$, $z_i$ being the elevation
of the mixed layer top.
(a): pdf's are shown without
any $r$-dependent rescaling; (b) the pdf is multiplied by the factor
$\sigma^{(z)} r^{-\zeta_\infty}$ with $\zeta_{\infty}\sim 0.6$:
the collapse of the curves indicate the asymptotic behavior
$P(\Delta_{r;z} \theta)\sim r^{\zeta_\infty}$ for large $\Delta_{r;z} \theta$,
that means saturation of temperature scaling exponents (see,
 and (\protect\ref{resca}), (\protect\ref{pawer}) and (\protect\ref{sat}));
(c) pdf's are multiplied by the factor $r^{\alpha^{(z)}}$
while $\Delta_{r;z} \theta$ by $r^{-\alpha^{(z)}}$:
the collapse of pdf cores indicates the validity of (\protect\ref{pdf-resc1})
that is equivalent to the
linear behavior of low-order  temperature scaling exponents 
(see (\protect\ref{mom})).

\item 
As in Fig.~2 but for $z/z_i=0.45$.

\item 
As in Fig.~2 but for $z/z_i=0.6$.

\item 
The cumulated probabilities 
$Prob[|\Delta_{r;z}\theta|>\lambda\sigma^{(z)}]$
for two values of $\lambda$
are shown for (a): $z/z_i=0.3$ and (b): $z/z_i=0.6$. 
The slope of these curves (continuous line) are compatible with the 
exponent $\zeta_{\infty}\sim 0.6$. The error bar on this slope is
of the order of  $0.1$, evaluated by means of the local scaling exponents (on
half-decade ratios) as customary in turbulence data analysis.

\item 
(a) Sixth and eight-order structure functions of temperature
differences 
vs the separation between points (squares). Stright lines have the
slope $\zeta_{\infty}=0.6$.  (b) The bulk contribution to the 
moments $p=6$ and $p=8$, $(\Delta_r \theta)^p P(\Delta_r \theta)$,
with $r/L\sim 1.1\times 10^{-1}$ in the inertial range of scales. 
For comparison $P(\Delta_r \theta)$
(i.e. for $p=0$) is also shown. Note that the maximum contribution
to the moments six and eight comes  from fluctuations $\Delta_r \theta$
in the region where $P(\Delta_r \theta)$ is well resolved. This proves
the reliability of our statistics to compute moments up to the order
eight.

\item 
The function $Q$ defined in (\protect\ref{resca}) is shown for two
different values of $z$ within the mixed layer: $z/z_i=0.3$
(dotted line) and $z/z_i=0.6$ (continuous line).
Differences in the shape of these two curves, reveal that $Q$ contains
a dependence on the elevation, $z$,  within the mixed layer.

\end{enumerate}

\newpage
\pagestyle{empty}

\begin{figure}
\vspace{-0cm}
\begin{center}
\includegraphics[scale=0.5,angle=0]{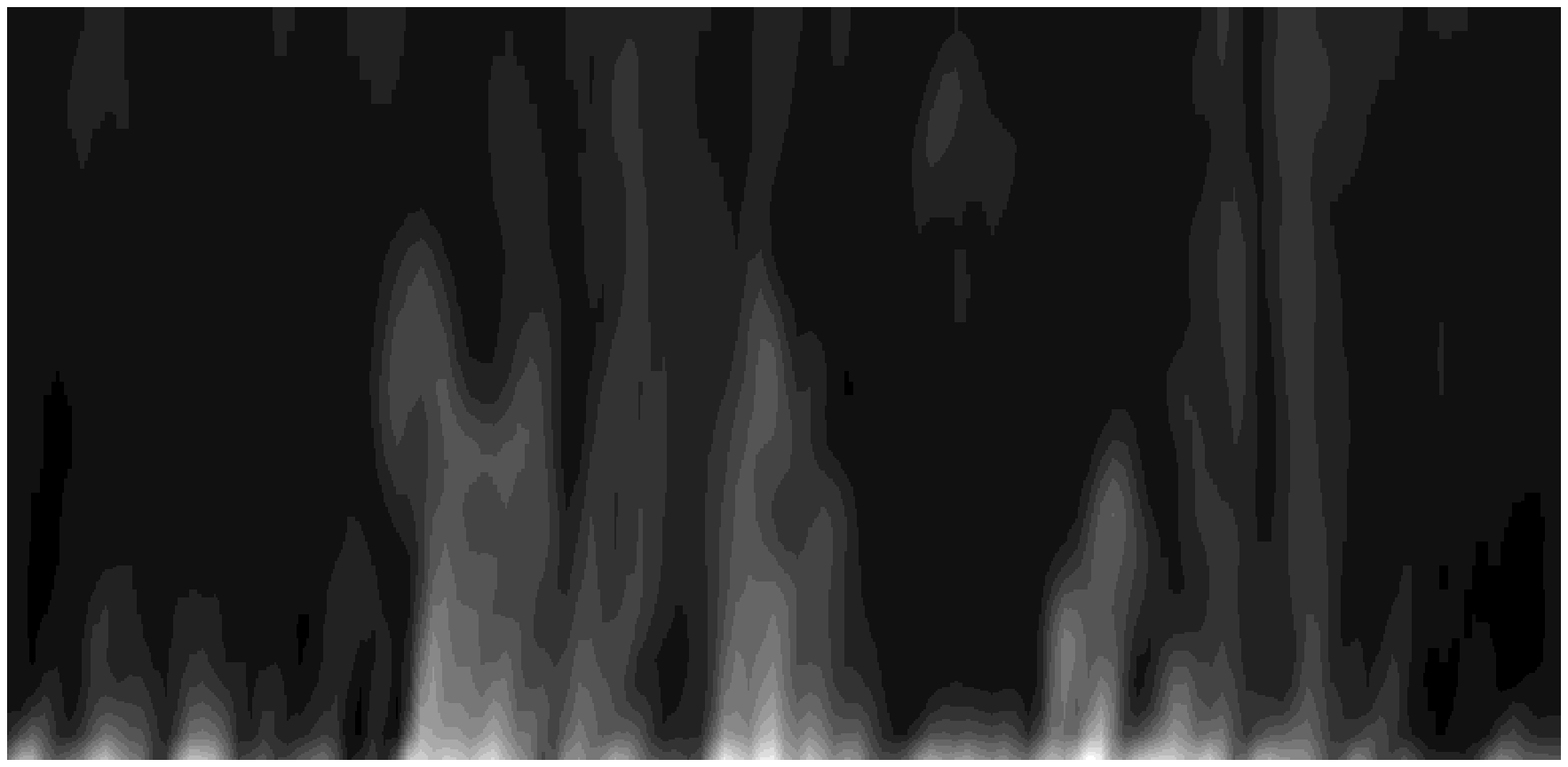}
\end{center}
\vspace{2cm}
\begin{center}
\hspace{2cm}\includegraphics[scale=0.5,angle=0]{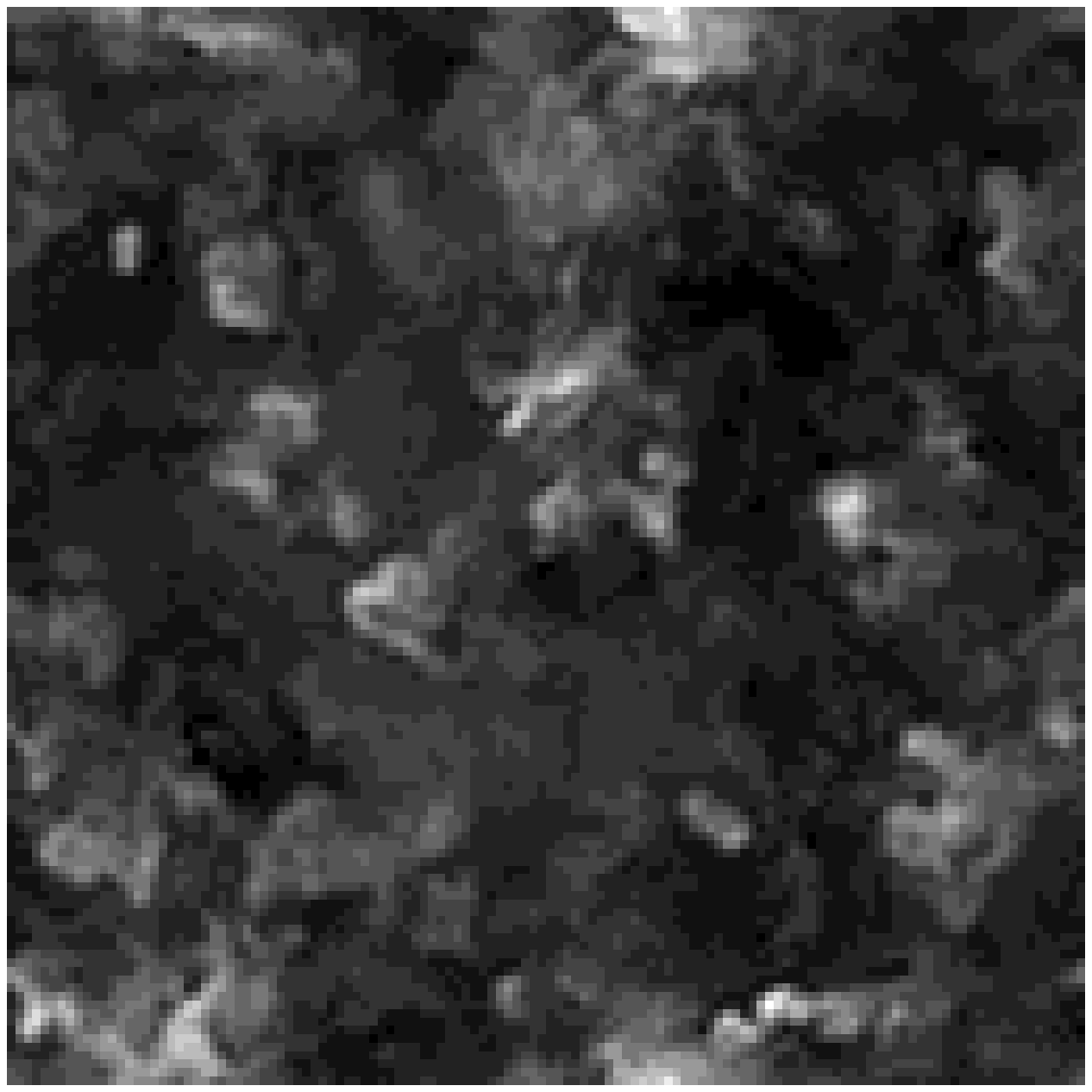}
\end{center}
\caption{}
\end{figure}
\begin{figure}
\vspace{-0cm}
\begin{center}
\includegraphics[scale=0.43,angle=-90]{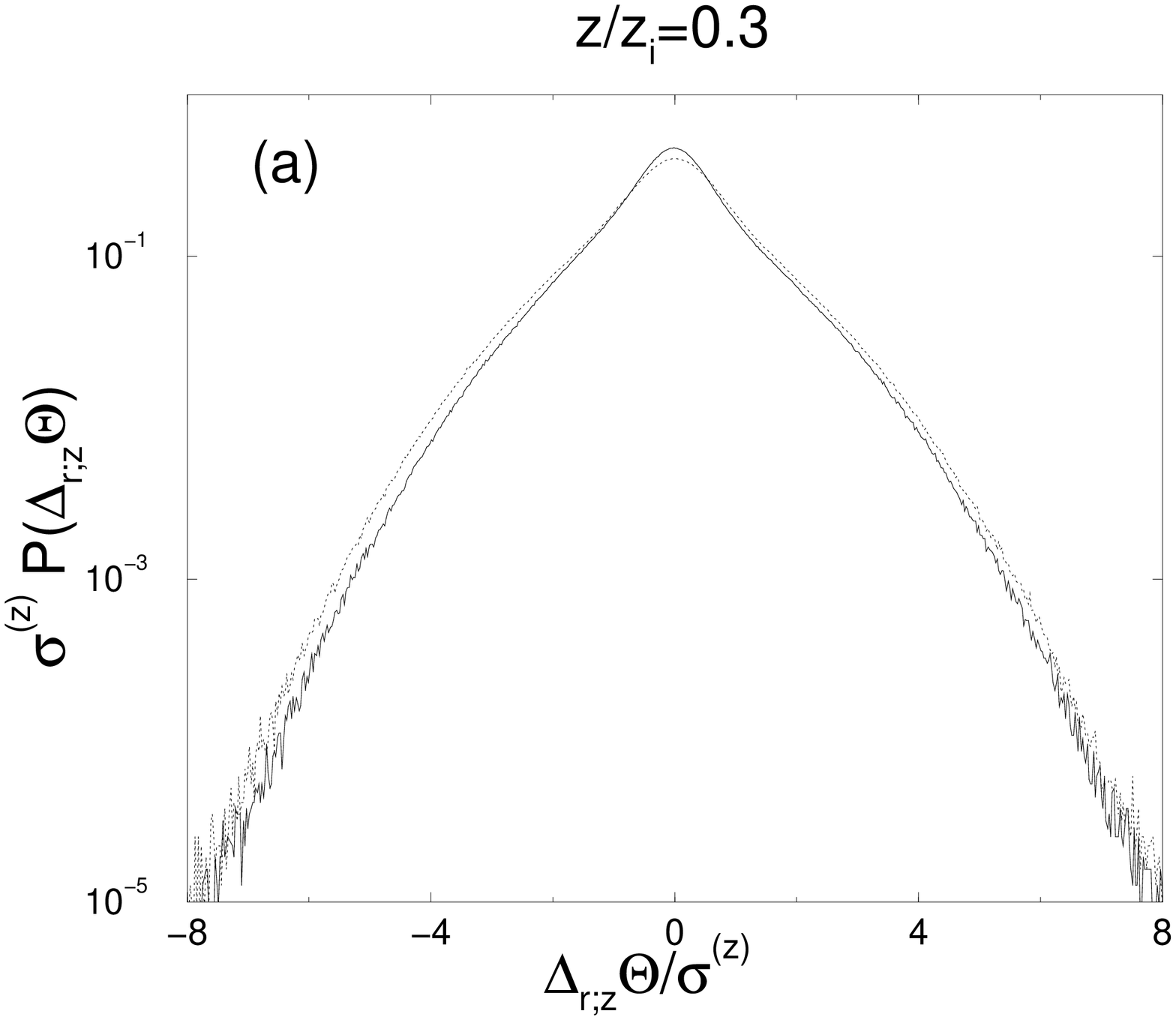}
\end{center}
\begin{center}
\includegraphics[scale=0.43,angle=-90]{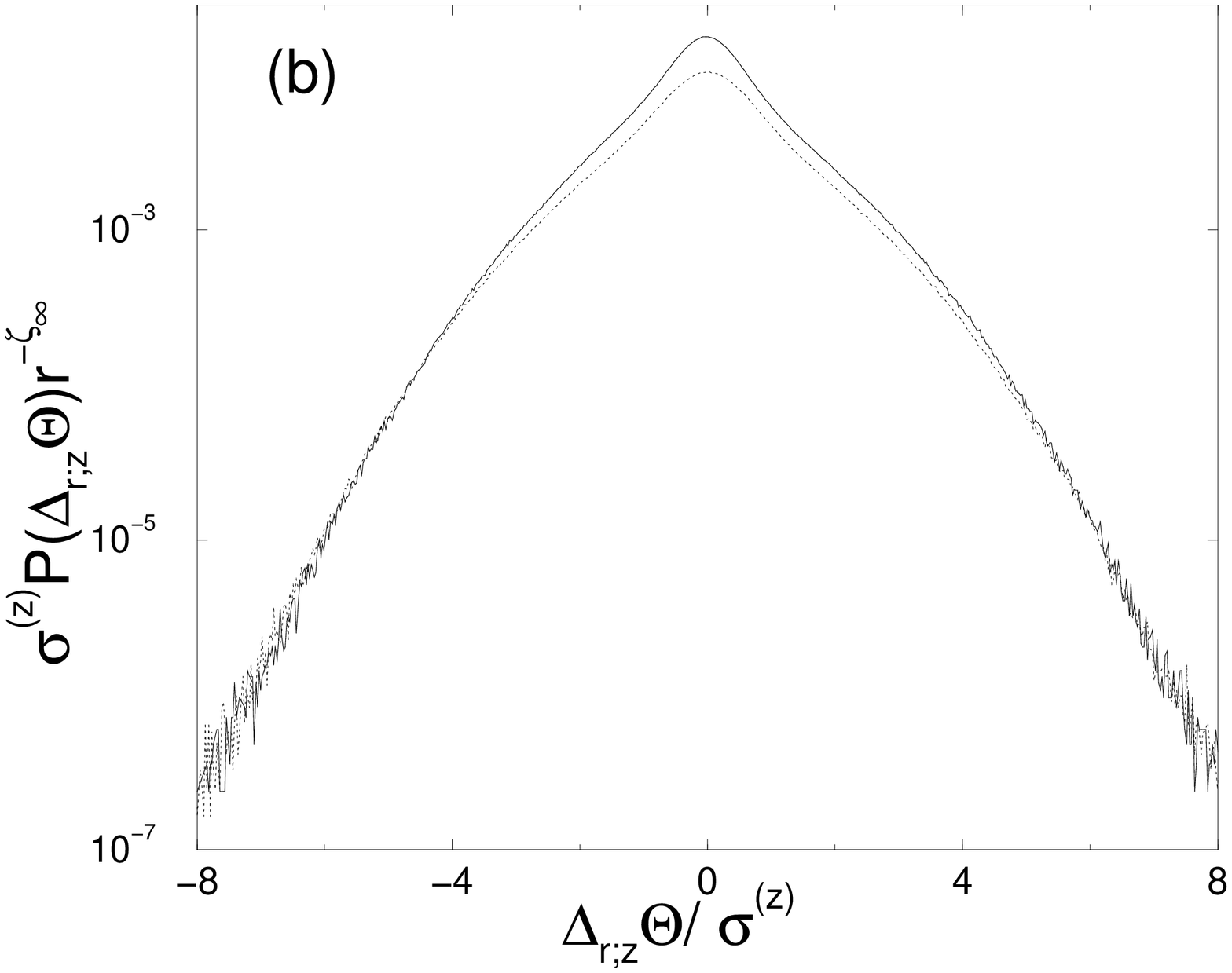}
\end{center}
\begin{center}
\includegraphics[scale=0.43,angle=-90]{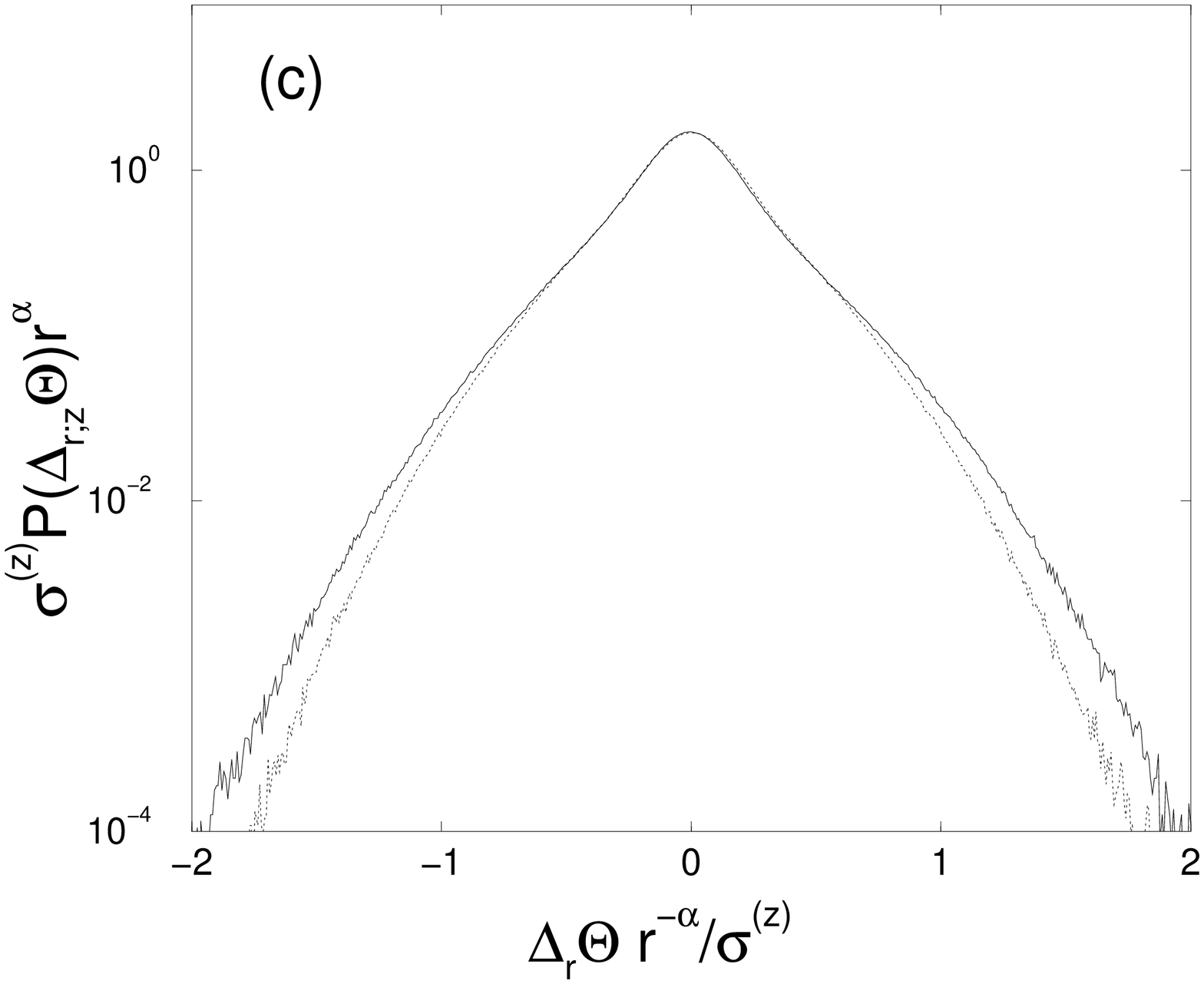}
\end{center}
\caption{}
\end{figure}
\begin{figure}
\vspace{-0cm}
\begin{center}\includegraphics[scale=0.43,angle=-90]{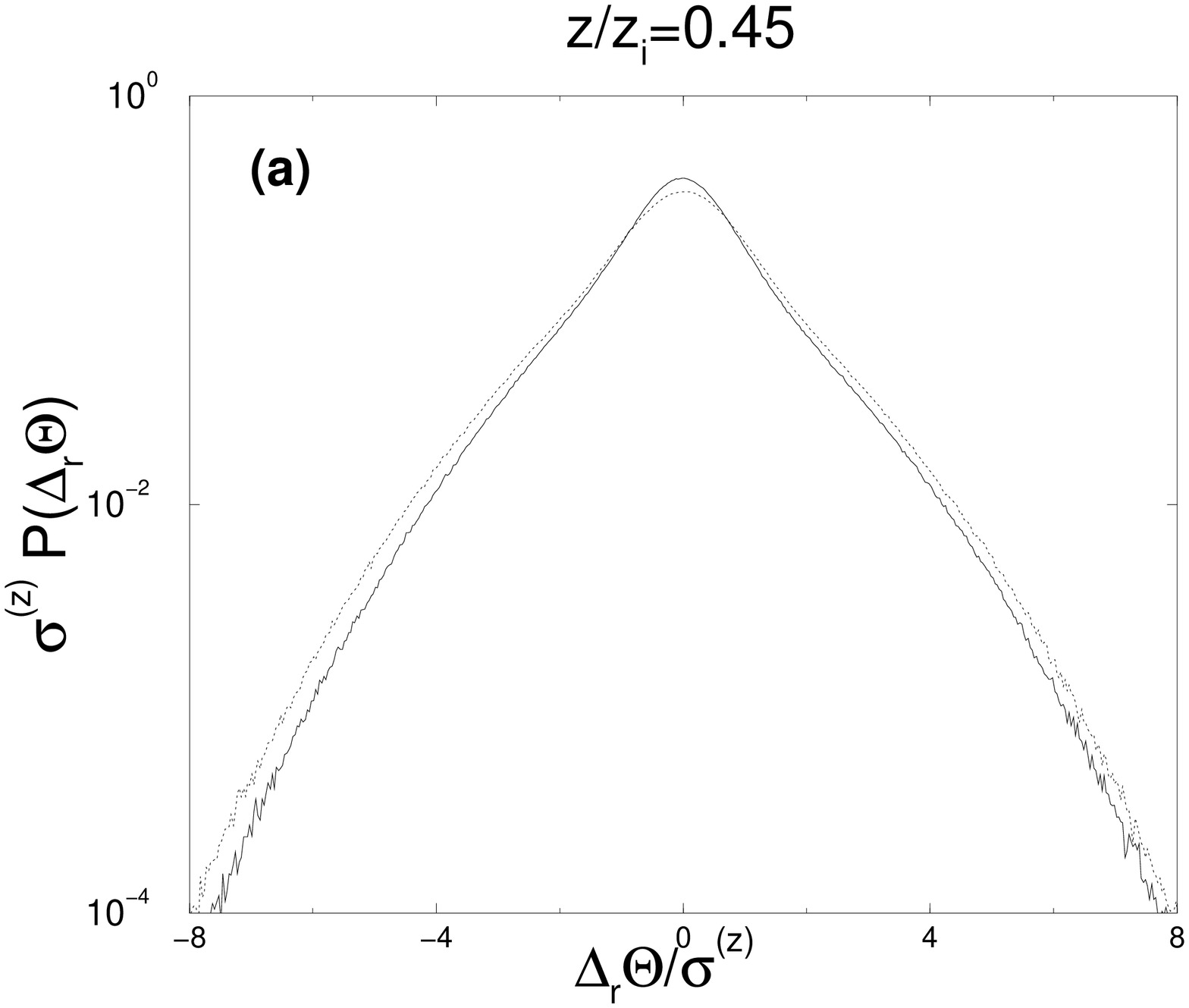}\end{center}
\begin{center}\includegraphics[scale=0.43,angle=-90]{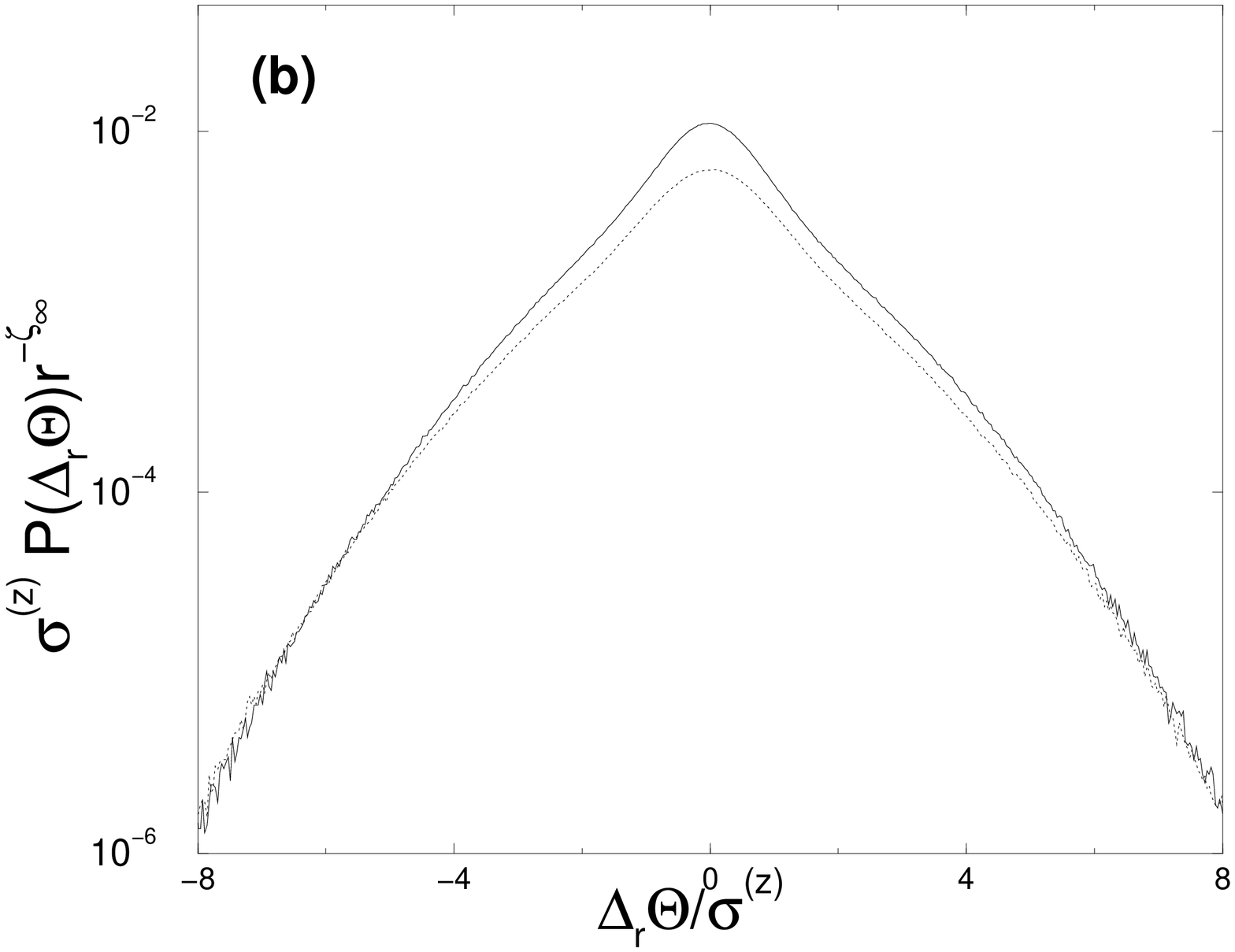}\end{center}
\begin{center}\includegraphics[scale=0.43,angle=-90]{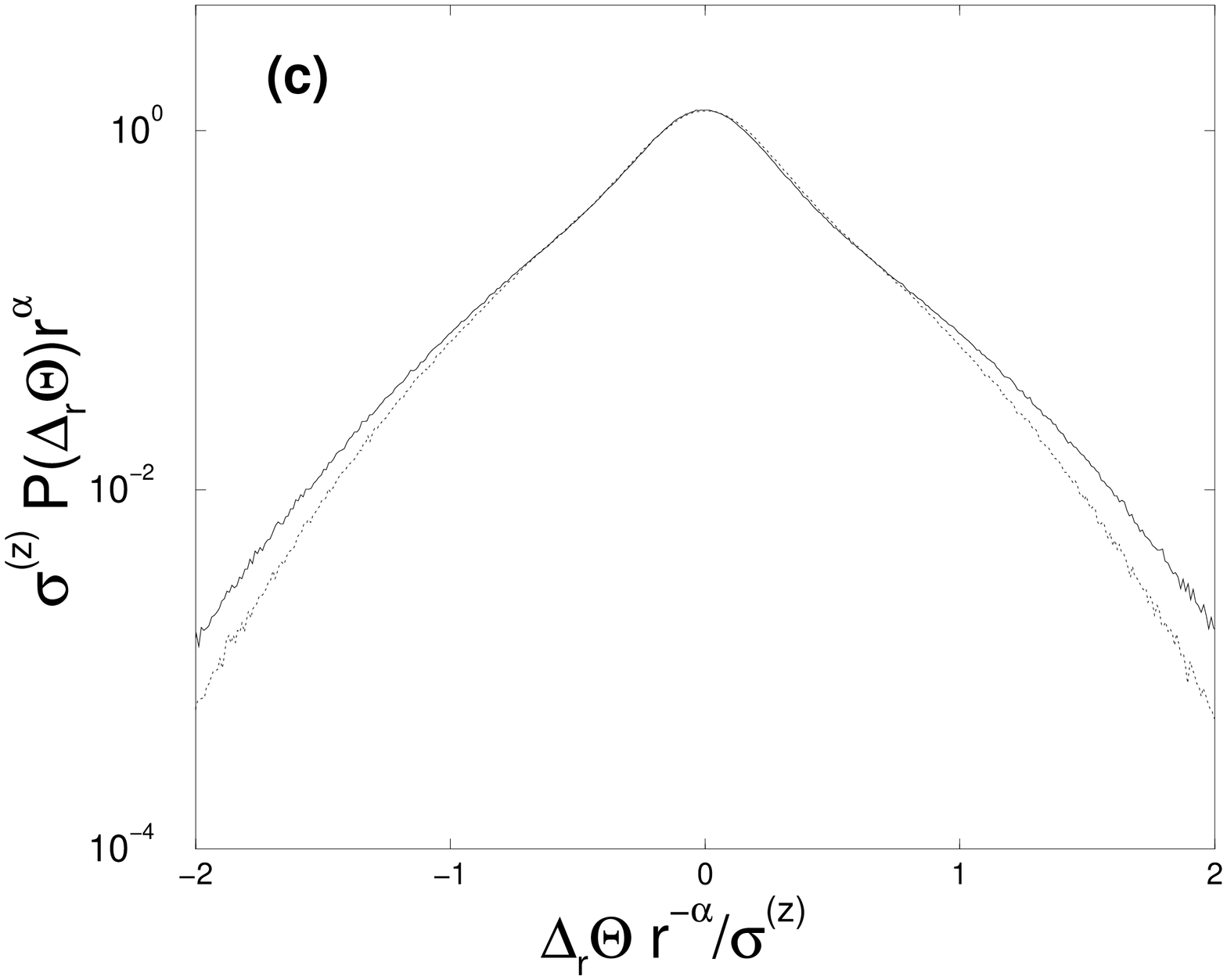}\end{center}
\caption{}
\end{figure}
\begin{figure}
\vspace{-0cm}
\begin{center}
\includegraphics[scale=0.43,angle=-90]{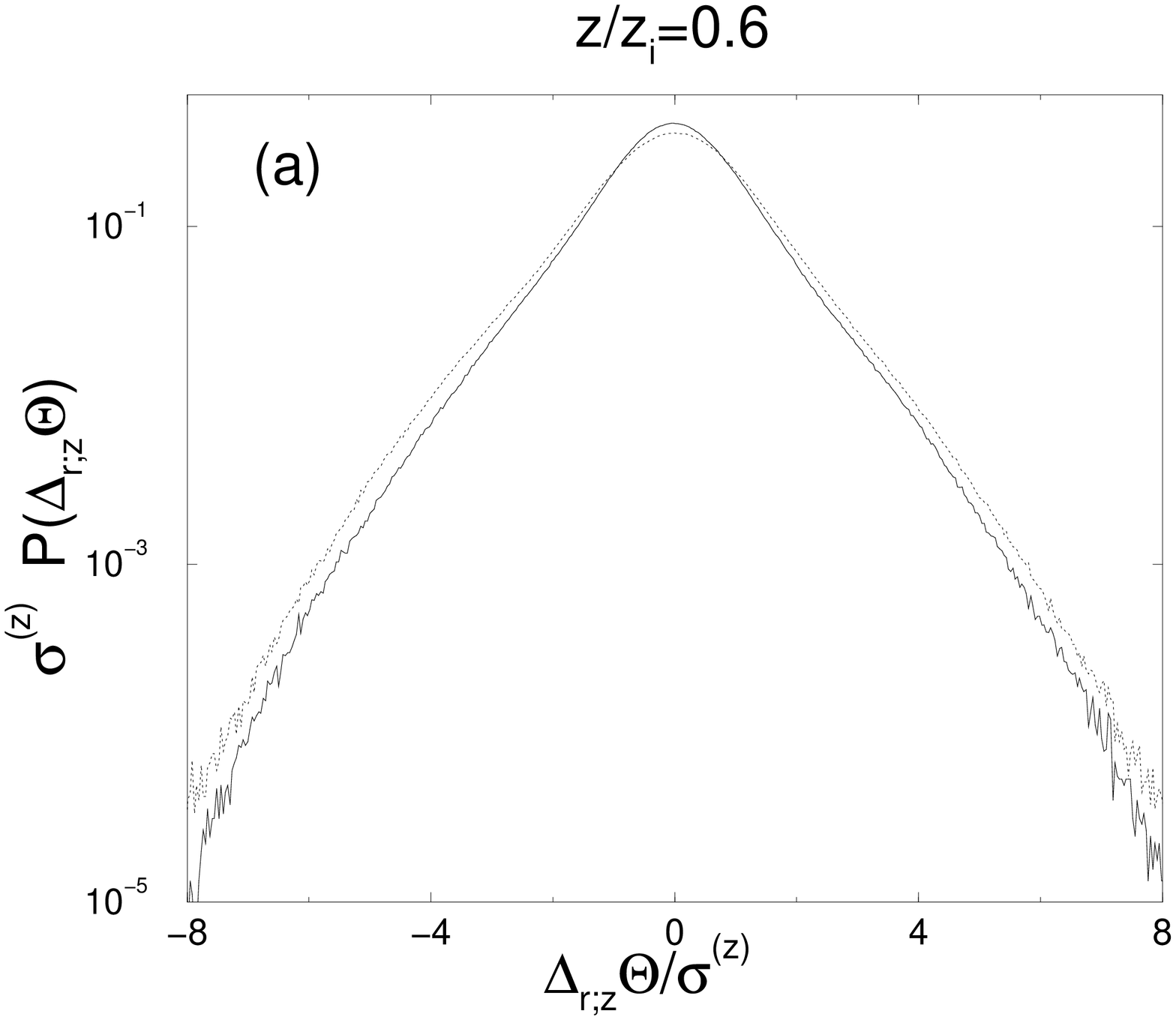}
\end{center}
\begin{center}
\includegraphics[scale=0.43,angle=-90]{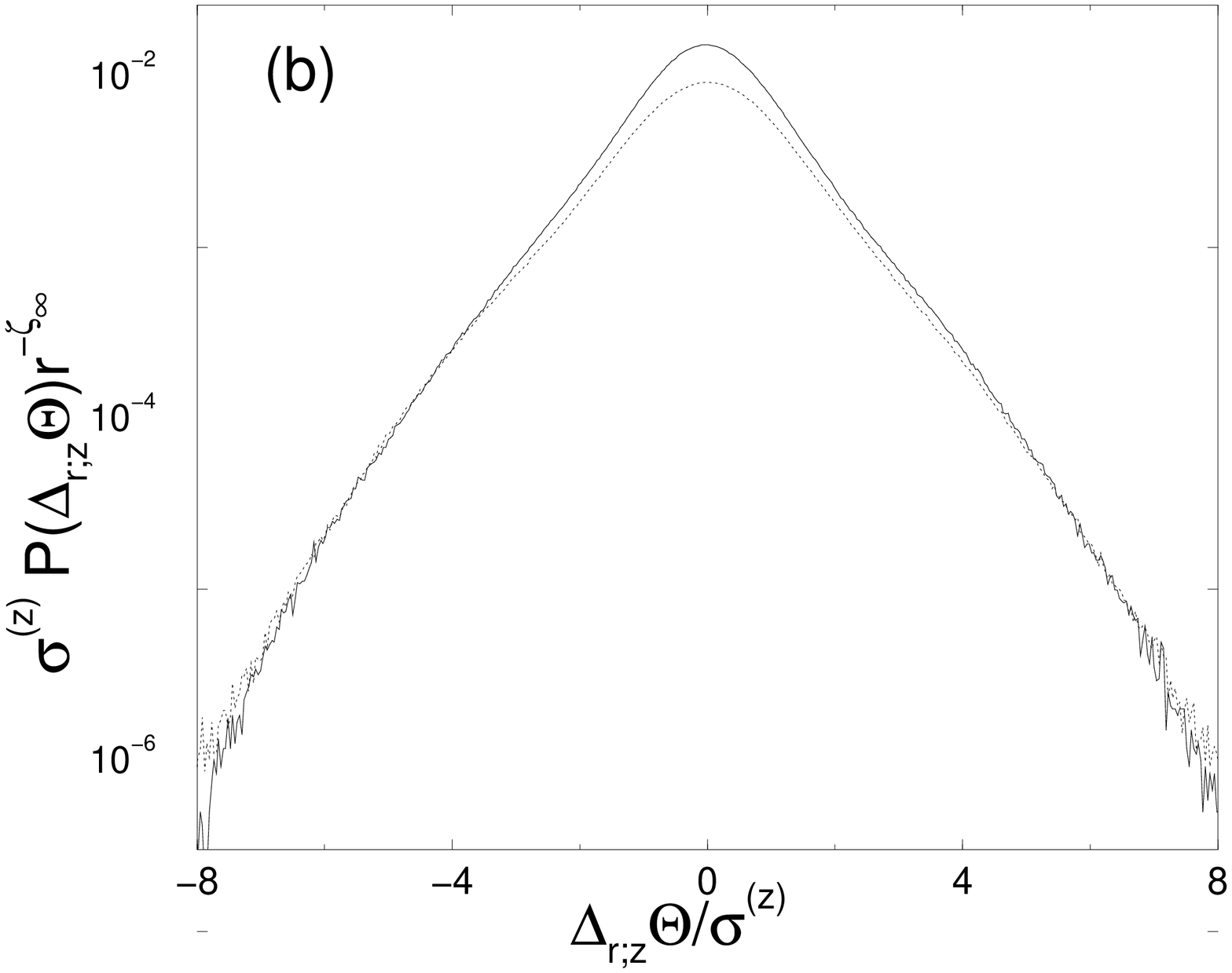}
\end{center}
\begin{center}
\includegraphics[scale=0.43,angle=-90]{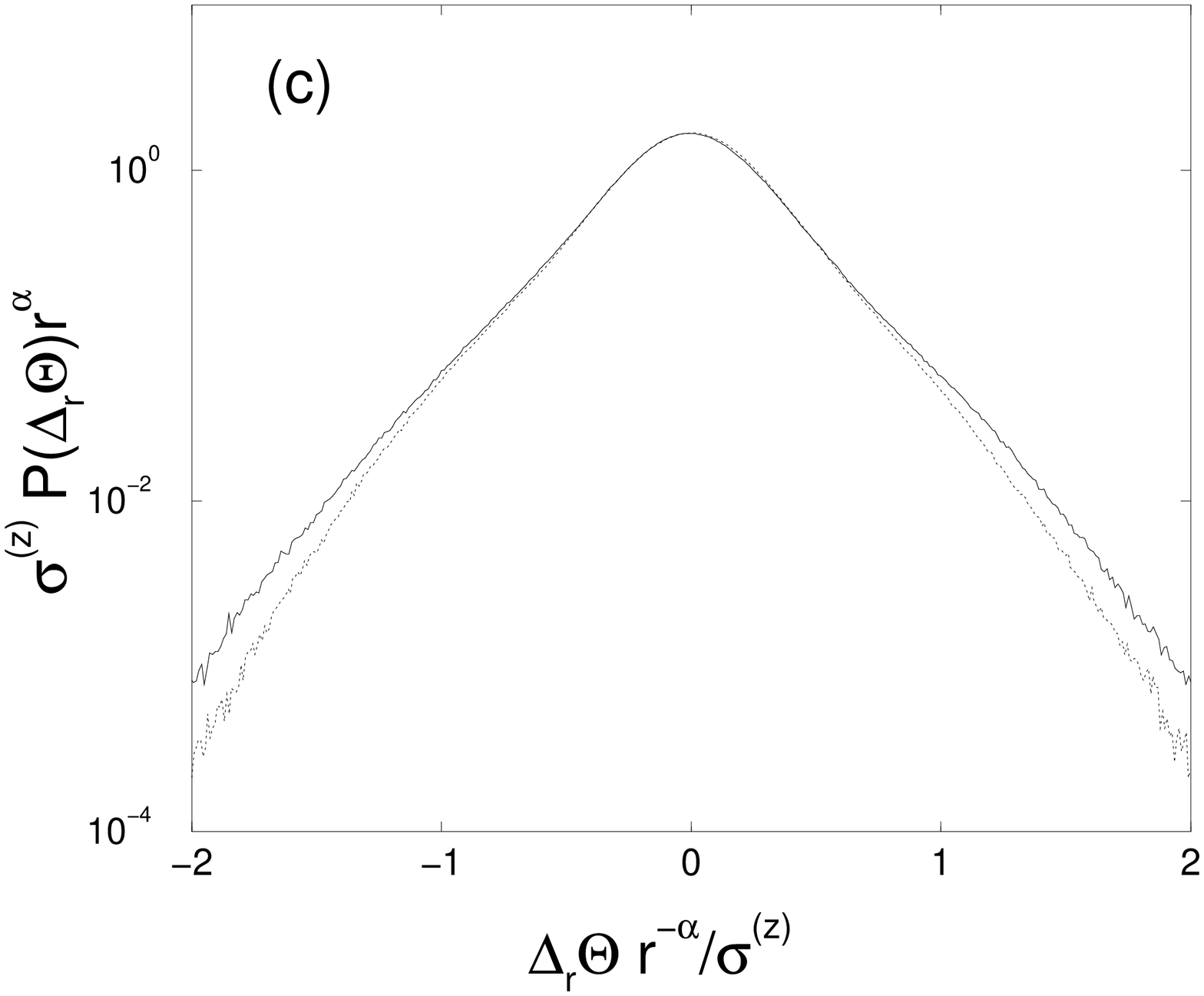}
\end{center}
\caption{}
\end{figure}
\begin{figure}
\vspace{-0cm}
\begin{center}
\includegraphics[scale=0.45,angle=-90]{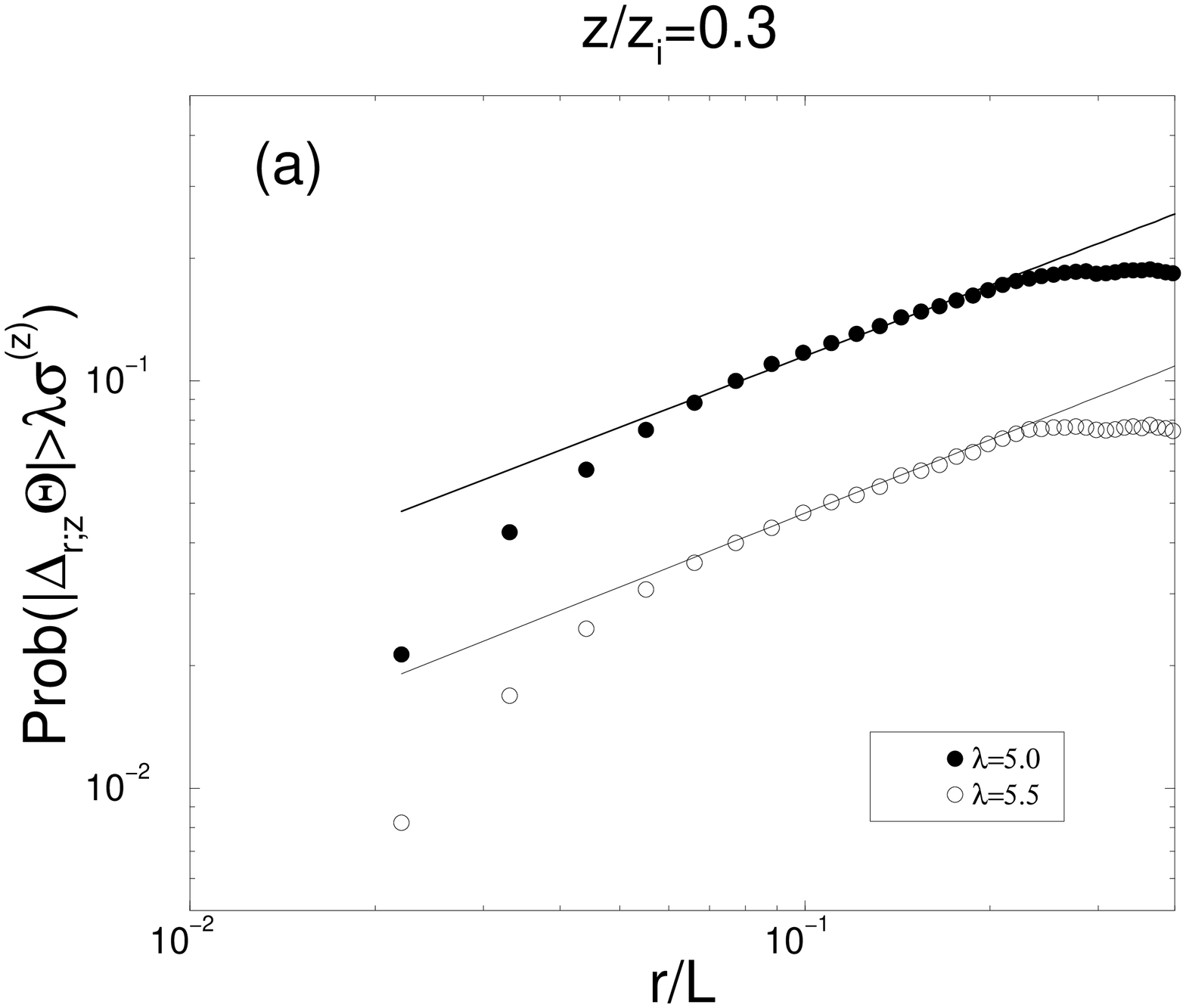}
\end{center}
\begin{center}
\includegraphics[scale=0.45,angle=-90]{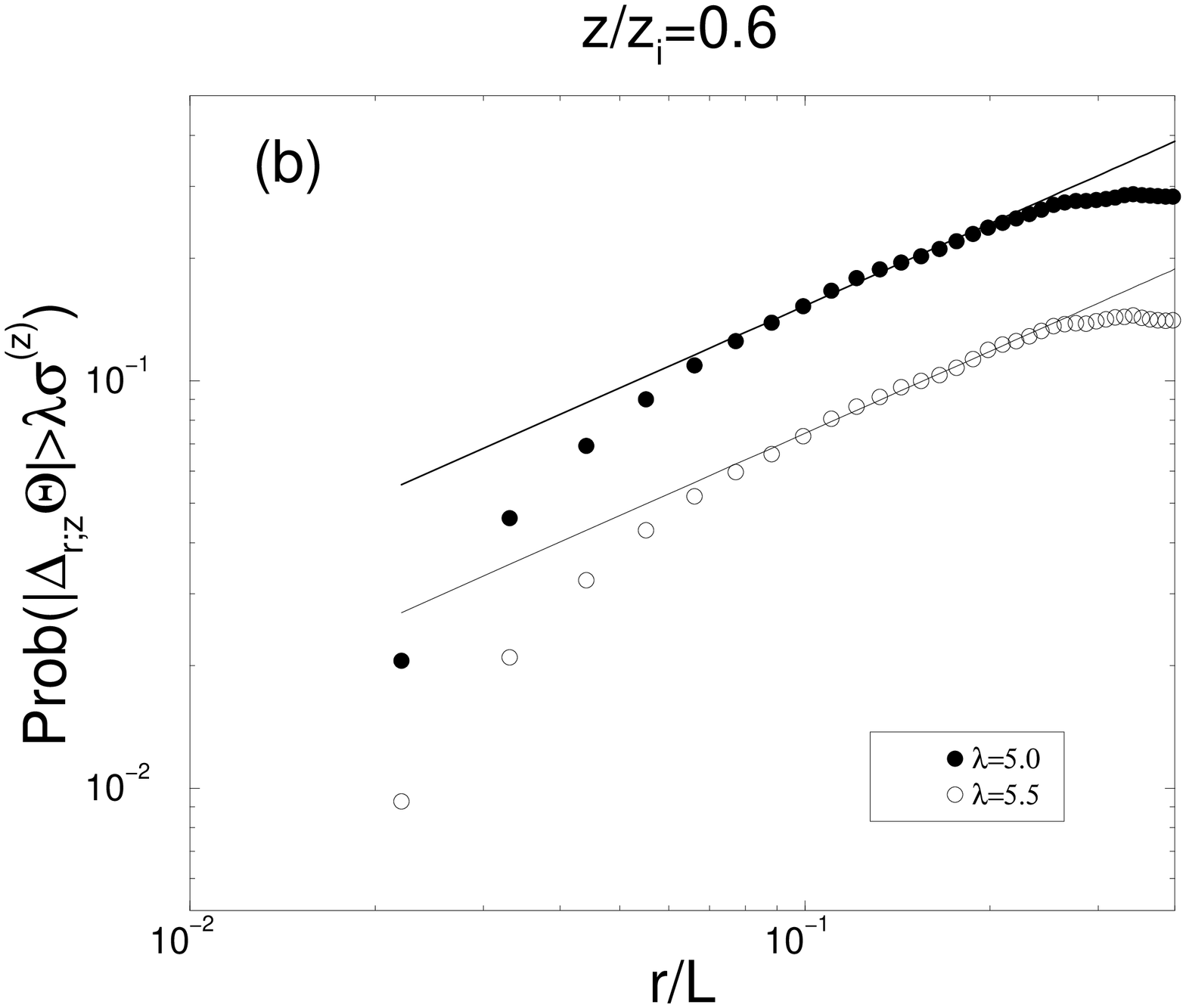}
\end{center}
\caption{}
\end{figure}

\begin{figure}
\vspace{-0cm}
\begin{center}
\includegraphics[scale=0.45,angle=-90]{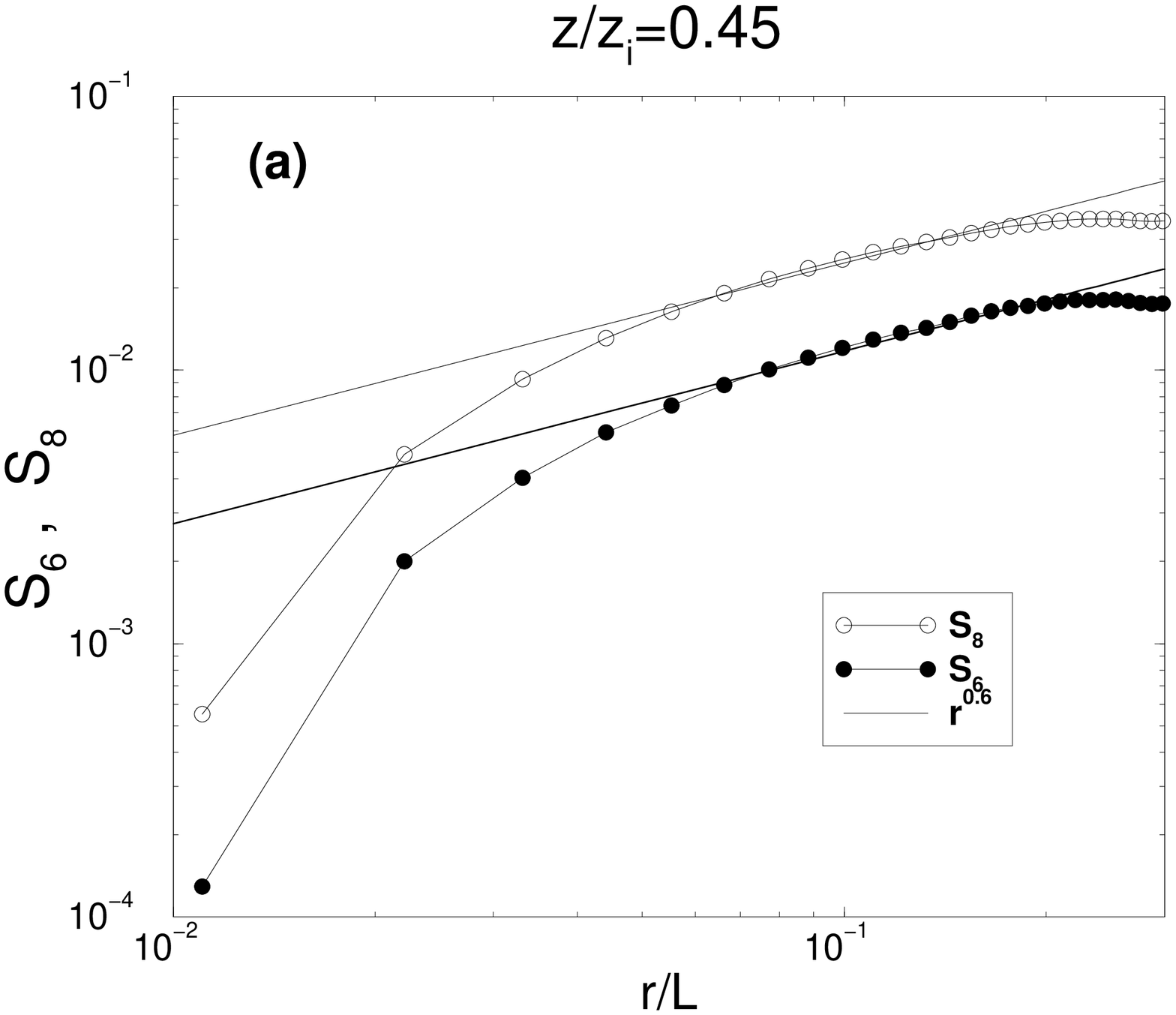}
\end{center}
\begin{center}
\includegraphics[scale=0.45,angle=-90]{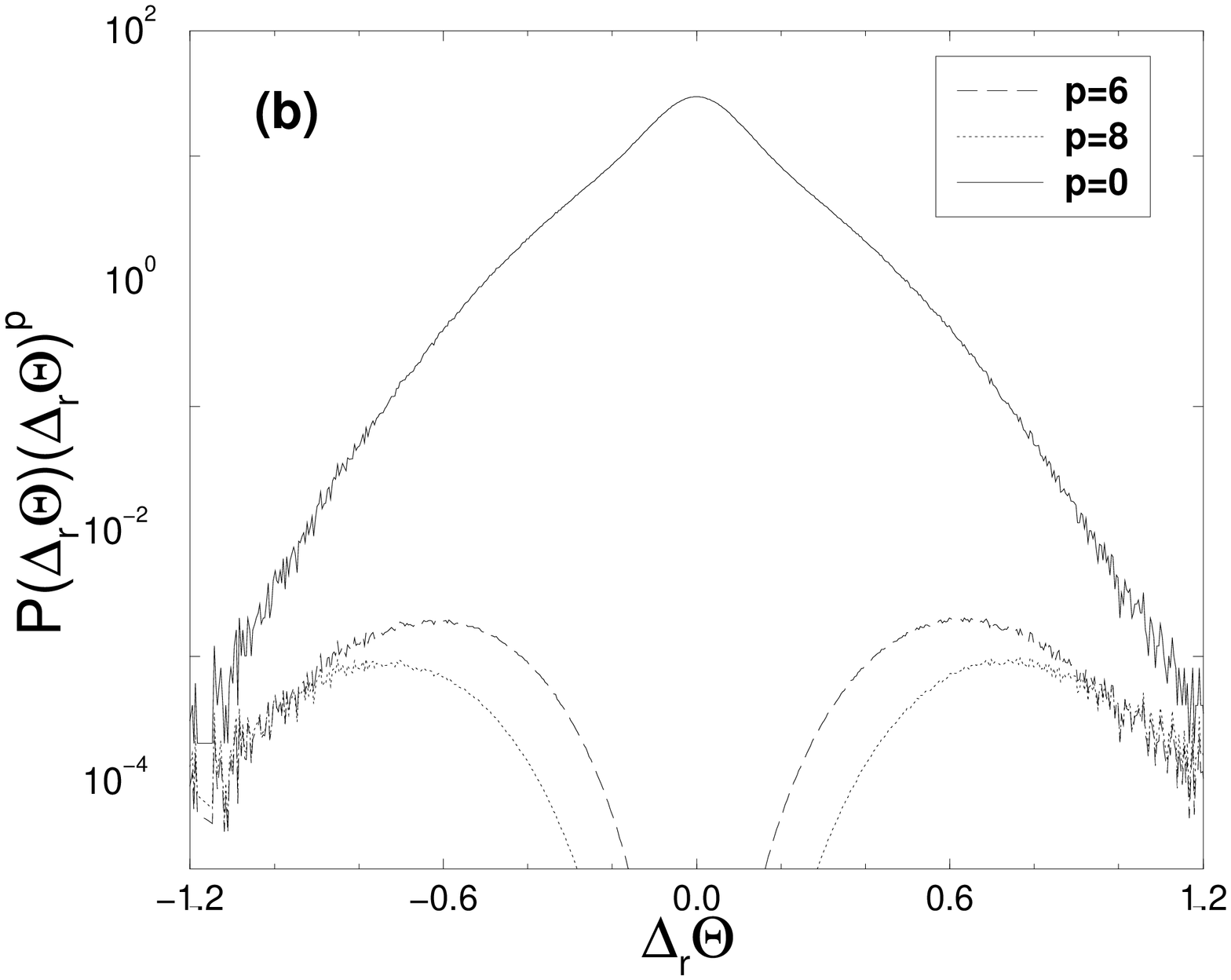}
\end{center}
\caption{}
\end{figure}
\begin{figure}
\vspace{-0cm}
\begin{center}
\includegraphics[scale=0.5,angle=-90]{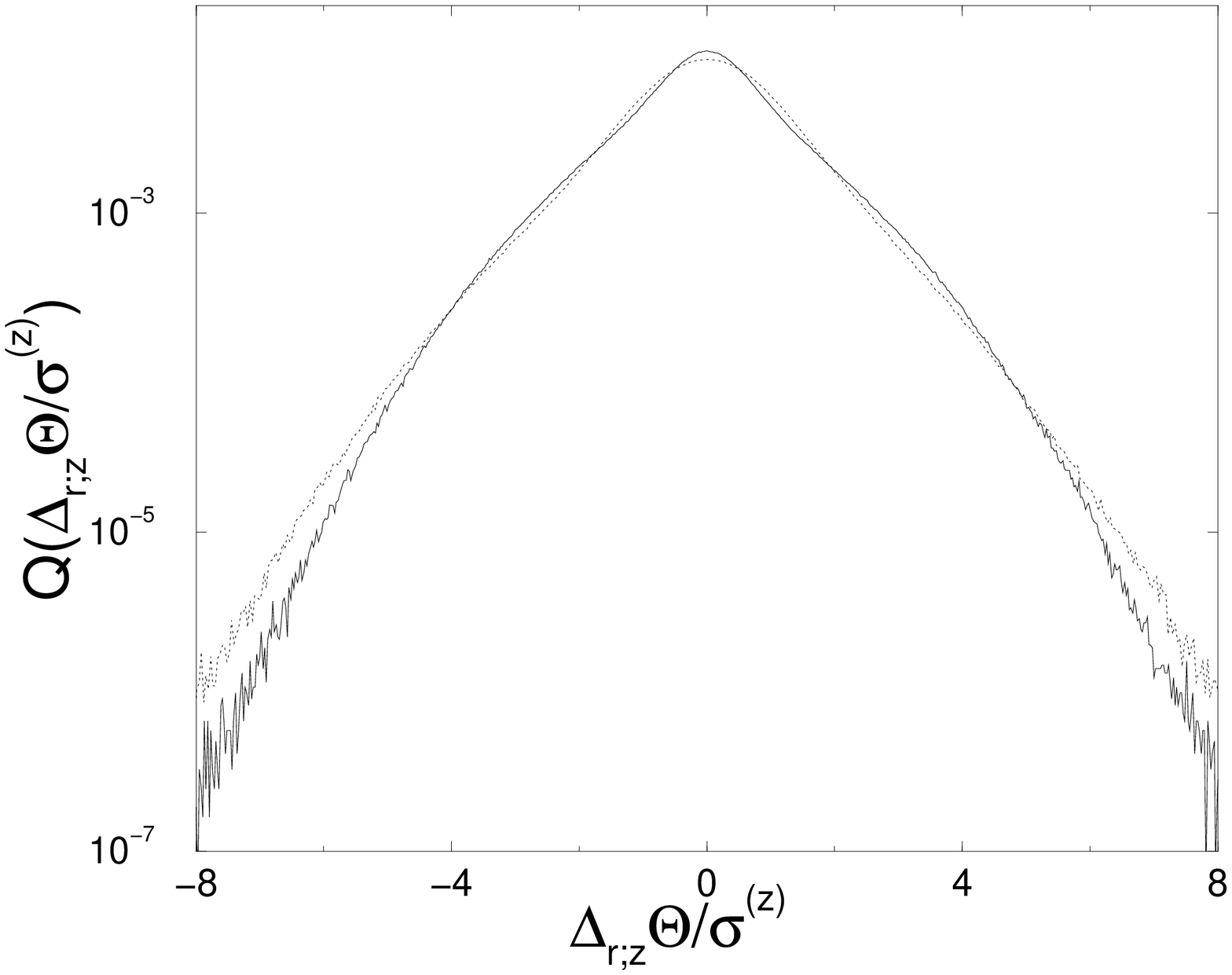}
\end{center}
\caption{}
\end{figure}

\end{document}